 \documentclass[final,5p,times,twocolumn,authoryear]{elsarticle}

\usepackage{amssymb}
\usepackage{amsmath}
\usepackage{lipsum}
\usepackage{csquotes}

\usepackage{fontenc}
\usepackage[utf8]{inputenc}
\usepackage{orcidlink}
\usepackage{xcolor}
\usepackage{booktabs}

\usepackage[normalem]{ulem}

\usepackage[mathscr]{euscript}
\usepackage[table]{xcolor}

\setcitestyle{numbers,square}

\begin{document}

\begin{frontmatter}




\title{Comparing the fusion of $^{19}$O and $^{20}$O ions with protons at near barrier energies}

\author[a]{Rohit Kumar\,\orcidlink{0000-0002-0450-7218}}
\author[a]{H. Desilets\,\orcidlink{0009-0009-4348-6526}}
\author[a]{R.~T. deSouza\,\orcidlink{0000-0001-5835-677X}}

\author[a]{S. Hudan\,\orcidlink{0000-0002-9722-2245}}

\author[b]{C. Ciampi\, \orcidlink{0000-0002-8142-0052}}
\author[b]{A. Chbihi\, \orcidlink{0000-0001-5653-4325}}

\author[c]{K.~W. Brown\, \orcidlink{0000-0003-1923-3595}}

\author[d]{B. Pinheiro\, \orcidlink{0000-0003-1025-3012}}

\author[d]{J. Lubian\, \orcidlink{0000-0001-7735-1835}}

\affiliation[a]{%
Department of Chemistry and Center for Exploration of Energy and Matter, Indiana University
2401 Milo B. Sampson Lane, Bloomington, Indiana 47408, USA}%

\affiliation[b]{GANIL, CEA,DRF-CNRS,IN2P3, 
Blvd. Henri Becquerel, F-14076, Caen, France} %

\affiliation[c]{Facility for Rare Isotope Beams and Department of Chemistry, Michigan State University, East Lansing, MI 48823, USA} %

 \affiliation[d]{organization={Instituto de F{\'\i}sica, Universidade Federal Fluminense},
             city={Niter\'{o}i},
             postcode={24210-340},
             state={RJ},
             country={Brazil}}

\begin{abstract}
Proton capture on $^{19,20}$O nuclei was measured in inverse kinematics with the active-target detector MuSIC@Indiana using CH$_4$ as the target gas. Direct measurement of the (p,n) cross section was enabled by the intrinsically high efficiency and self-normalizing nature of the measurement. The excitation function of both (p,n) reactions exhibits an oscillatory behavior. For the neutron-rich nuclei studied, determining the (p,n) cross section provides access to the proton fusion cross section because the proton's de Broglie wavelength is much larger than that of the heavy ion.  A significant suppression of the cross section in $^{19}$O compared to $^{20}$O is observed, suggesting level-density differences. While the coupled channel calculations reproduce the oscillatory behavior observed in the case of $^{20}$O, they fail to predict the oscillatory behavior observed in $^{19}$O or the overall suppression of the fusion cross section. 

\end{abstract}




\end{frontmatter}




\section{Introduction}
\label{introduction}

Neutron and proton capture reactions have played a central role in nuclear physics since its inception and remain a topic of both fundamental and practical interest. Observation of neutron resonances following neutron capture on heavy nuclei was instrumental in realizing that the decay of these nuclei could be treated in a semiclassical fashion \cite{Anderson50, Fenstermacher59, Kennett58}. Key to this treatment is the concept of the level density parameter \cite{Hurwitz51, Gadioli68, Greaves25}, intrinsic to a Hauser-Feshbach description of the statistical decay of an excited nucleus.  Proton capture on light nuclei is also interesting, as these reactions are responsible for synthesizing the lightest isotopes of an element in astrophysical environments \cite{Arnould03}. 

Proton and neutron capture reactions are also interesting as a reference for heavy-ion fusion.
They represent the situation of minimal Coulomb repulsion and minimum angular momentum. Minimizing Coulomb excitation of the two approaching nuclei minimizes the impact of low-lying collective modes on the fusion cross-section, simplifying its description.  
At the incident energies considered in this work, proton capture is synonymous with proton fusion because the proton's de Broglie wavelength is much larger than that of a heavy ion.

Fusion of a neutron-rich incident nucleus with a proton at the incident energies investigated results in a compound nucleus with an excitation of $\sim$12 MeV, above the neutron separation energy. In general, the compound nucleus de-excites by neutron, charged-particle, and $\gamma$ emission. However, because the neutron separation energy is relatively low compared with that of charged particles, single-neutron emission dominates the de-excitation to the evaporation residue. The probability of forming the evaporation residue is governed by both the probability of forming the compound nucleus and the probability of its decay. The former depends on the level density of the quasi-bound system as the two nuclei merge \cite{Desilets26}, while the latter depends on the level density of the evaporation residue.

Recently, an exciting result for proton capture by $^{20}$O was reported \cite{Desilets26}. An oscillatory behavior of the fusion excitation function at energies just above the barrier was observed. These oscillations, with widths of 50-- 100 keV, were interpreted as evidence for quasi-bound states in the fused system. 
In this work, we examine proton capture by $^{19}$O and determine if it also manifests an oscillatory behavior in its excitation function. Moreover, a quantitative comparison between the proton capture excitation functions of $^{20}$O and $^{19}$O is made, and both excitation functions are compared with the predictions of coupled channel calculations.


\section{Experimental details}

The radioactive beam experiment was conducted at the SPIRAL1 facility of the GANIL accelerator complex in Caen, France. 
The $^{19}$O (t$_\frac{1}{2}$ = 26.5 s) and $^{20}$O (t$_\frac{1}{2}$ = 13.5 s) beams were produced by bombarding a graphite target with a  $^{22}$Ne primary beam at E/A = 80 MeV. The  $^{19}$O (or $^{20}$O) ions produced were then accelerated by the CIME cyclotron to an energy of E/A = 3.0 MeV (2.7 MeV) and selected in B$\rho$ by the ALPHA spectrometer before being transported to the experimental setup.

Multi-Sampling Ionization Chamber (MuSIC) detectors are a powerful tool in investigating nuclear reactions with radioactive beams \cite{Johnstone21, Carnelli15, Blankstein23, Asher21b}. These detectors have been used to study ($\alpha$, p) \cite{Avila17, Jayatissa22} and ($\alpha$,n) \cite{Avila17, Blankstein24, Ong22} reactions of astrophysical interest as well as heavy-ion fusion \cite{Desilets25a, Hudan24, Carnelli14, Asher21a}. 

The key element in the experimental setup was the active-target detector, MuSIC@Indiana \cite{Johnstone21}. 
This detector is a transverse-field ionization chamber with a segmented anode. The anode consists of twenty 12.5 mm-wide strips 
oriented perpendicular to the beam axis, thus allowing the ionization of each incident beam ion or heavily ionizing reaction product to be tracked as it traverses the gas. The energy deposited on each anode strip is measured by integrating the charge deposited on each anode using 
high-quality charge-sensitive amplifiers \cite{zepto}, shaping amplifiers, and peak-sensing Caen V785 ADCs. Digitized data is recorded by means of a standard VM-USB \cite{Wiener} PC-based data acquisition system.

In the present experiment, the detector was filled with high-purity CH$_{4}$ at pressures between 110 and 130 torr.  
All events with incident ions depositing at least 1.0 MeV in the detector were recorded for subsequent analysis.
Detailed descriptions of the detector design and performance can be found in Refs.~\cite{Johnstone21,Desilets25}. In the present experiment, the beam was incident on the detector at an intensity up to $\sim$1-2$\times$10$^4$ ions/s.

For each incident ion, the set of measured anode-strip energies, E$_{\rm A}$, is collectively referred to as a \enquote{trace}.
Traces associated with proton capture are readily
distinguished from the beam by a sudden, but modest increase in ionization relative to the beam \cite{Desilets26,Johnstone21}. This increased ionization is observed over several anodes and is consistent with proton capture/fusion.
Proton fusion traces are easily distinguished from those associated with carbon fusion, as the latter manifest a much larger increase in energy deposit due to the significantly larger increase in atomic number for the fusion product.
Two-body processes (inelastic scattering and transfer) exhibit an entirely different energy deposition in the detector. 
The ability to distinguish these three reaction types with MuSIC@Indiana has been previously described \cite{Desilets26, Johnstone21,Hudan24}.

Fusion of a proton with an $^{19,20}$O ion results in the formation of an excited $^{20,21}$F compound nucleus with an excitation energy of 12.1 MeV $\leq$ E$^*$ $\leq$ 13.3 MeV.
The low neutron binding energy (6.6 MeV and 8.1 MeV for $^{20}$F and $^{21}$F, respectively), compared with that of a proton or an $\alpha$-particle, leads to de-excitation by neutron emission to $^{19,20}$F.  By distinguishing the resulting F from the incident O ions, measurement of the (p,n) cross section is accomplished. It should be emphasized that the inverse kinematics of the measurement together with the high efficiency and self-normalizing character of the MuSIC approach makes this a highly effective means to measure the (p,n) cross section.

\section{Analysis} 
The challenge faced in isolating proton fusion with radioactive beams using an active target detector is the modest separation of the reaction product with $\Delta$Z = 1 from the beam. This challenge is similar to the one confronted in investigating ($\alpha$,p) reactions \cite{Jayatissa22} and is somewhat more challenging than studying ($\alpha$,n) reactions \cite{Avila17, Blankstein24}, which have $\Delta$Z = 2.

\begin{figure}[h]
\begin{center}
\includegraphics[width=0.35\textwidth]{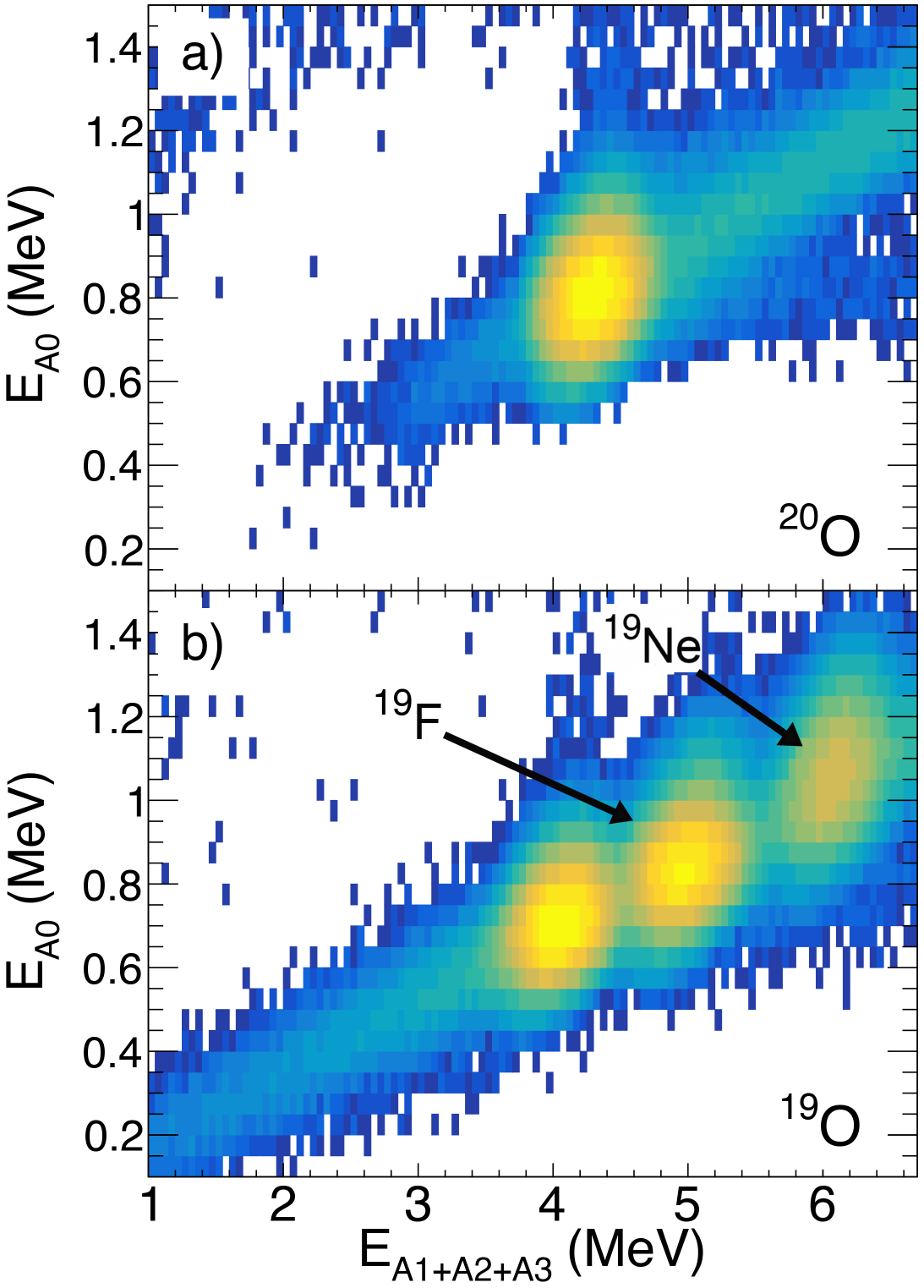}
\caption{Identification of the incident beam based upon the energy deposit in the first four anodes of MuSIC@Indiana. Panel a) For the incident $^{20}$O beam, no evidence of beam contamination is observed. Panel b) Three peaks are observed corresponding to the presence of $^{19}$F ($\sim$37\%) and $^{19}$Ne ($\sim$4\%) along with $^{19}$O beam. However, when averaging over all the measured dataset $^{19}$F is $\sim$55\% and $^{19}$Ne remains unchanged.}
\label{fig:PID}
\end{center}
\end{figure}

Identification of the incident beam was accomplished by utilizing the first four anode strips of MuSIC@Indiana for an E$_{A0}$ vs E$_{(A1+A2+A3)}$ measurement as indicated in Fig.~\ref{fig:PID}. In Fig.~\ref{fig:PID}a, for $^{20}$O, a single dominant peak is observed. The locus extending to higher energy deposits from this peak is associated with pileup events in the detector, which occur at the 1$\%$ level.  
In Fig.~\ref{fig:PID}b in contrast to the case of $^{20}$O,
three peaks are observed for $^{19}$O. These peaks indicate the presence of $^{19}$F and $^{19}$Ne contaminants at a level of $\sim$37\% and $\sim$4\% respectively. Both contaminants are well separated from the $^{19}$O beam of interest.  Incident  
O ions are selected by gating on the appropriate peak in the two-dimensional spectrum providing rejection of beam contaminants, any reactions arising from the mylar entrance window, and pileup events. 

To focus on events in which the incident ion had either reacted or scattered in the detector, the outgoing beam was rejected by examining the characteristic energy loss measured in E$_{A18}$ vs E$_{A19}$.
Following this selection, residual events corresponding to the beam are further rejected by requiring E$_{A}^{max}$ $>$ E$_{th,min}$, where
E$_{A}^{max}$ is the maximum anode energy deposited within a trace. Fusion of the incident beam with carbon is rejected by requiring that E$_{A}^{max}$ $<$ E$_{th,max}$. 

\begin{figure}[h]
\begin{center}
\includegraphics[width=0.45\textwidth]{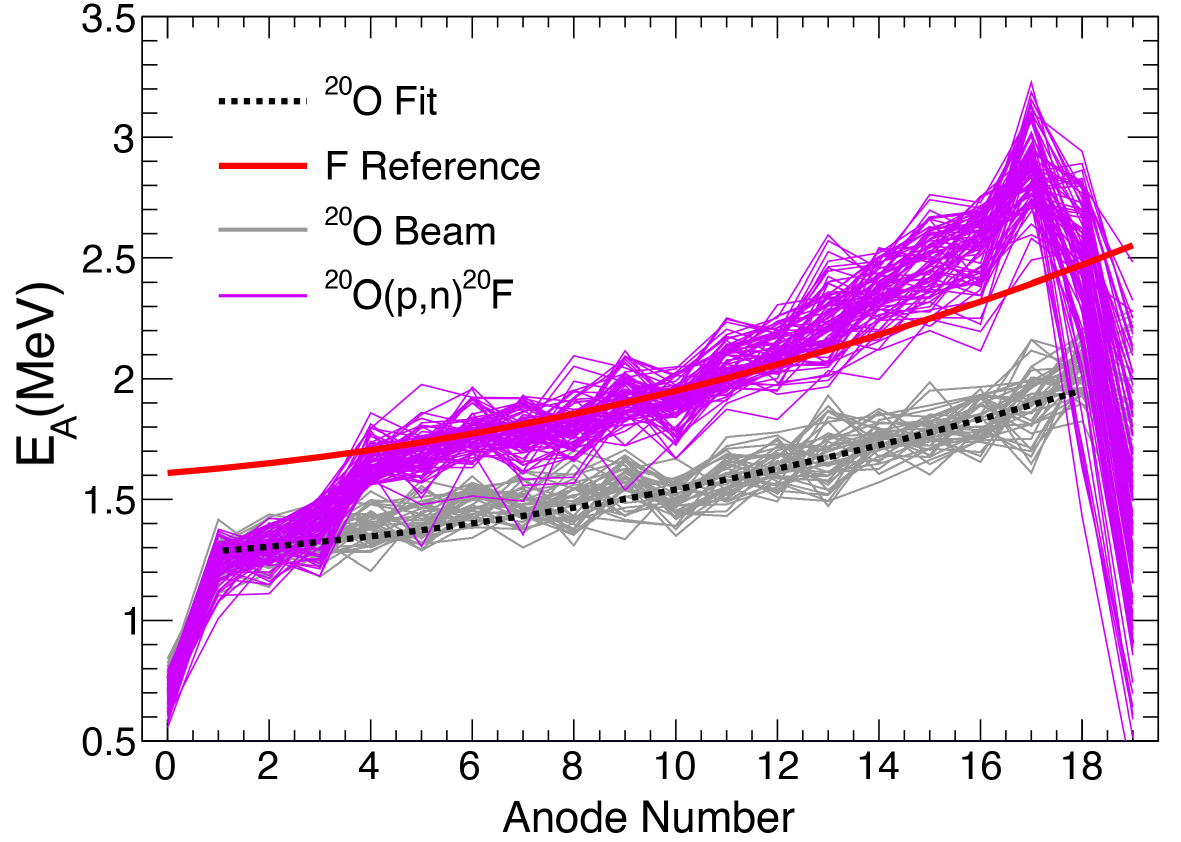}
\caption{Putative proton fusion trace (solid, magenta) is compared to a few beam traces (solid, gray). The average of the beam traces, shown as the dotted (black) line, indicates the reference beam trace. Indicated by the solid (red) line, based upon a 
Z$^2$ scaling is the reference line for F ions.}
\label{fig:OFAvg}
\end{center}
\end{figure}

Events were subsequently designated as more \enquote{O-like} or \enquote{F-like}.
This classification was realized by calculating the deviation of a given trace from O and F reference traces. The reference trace for O was calculated by fitting the traces of representative beam events, defined by the E$_{A18}$ vs E$_{A19}$ selection, with a second-order polynomial. 
Fig.~\ref{fig:OFAvg} shows representative beam traces along with the reference line for O. 
To calculate the reference line for F, the Z$^2$ dependence of the energy deposit for ions of the same velocity was assumed, consistent with the Bethe-Bloch formula \cite{Bethe30, Bloch33}.
The reference F line, along with candidate proton fusion traces that manifest increased ionization after anode 4 and have maximum ionization in anode 17, are also presented in Fig.~\ref{fig:OFAvg}.

The reasonably good agreement of the traces to the F reference for several anodes after their initial increase from beam ionization suggests that these traces correspond to F ions.  For these candidate proton fusion events, the resulting ER reaches a maximum energy deposit in anode 17 before stopping in the detector volume. As expected, near the Bragg maximum, Z$^2$ scaling is a poor description, likely because of changes in the ion charge state.

For a given trace, the deviation from the F reference line, $D_F$, was calculated as:
\begin{equation}
D_F=\sum_{i=A_{th}}^{A_{max}}{(E_A(i)-E_{A, Fref}(i))^2}
\end{equation}

where $A_{th}$ is the first anode whose energy deposit is 150 keV above the beam average. The summation runs up to the anode with the maximum energy deposit in the trace, $A_{max}$. The deviation from the O reference, D$_O$, was calculated analogously. 

\begin{figure}[h]
\begin{center}
\includegraphics[width=0.45\textwidth]{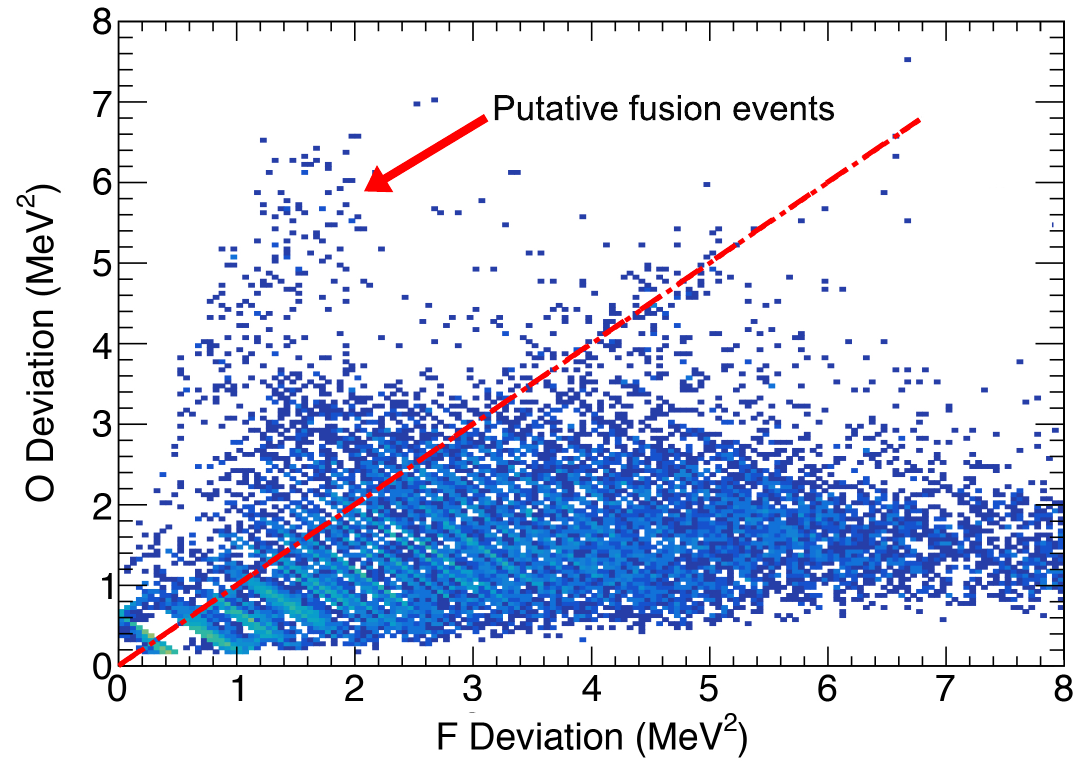}
\caption{Deviation of each event trace from the average O and F reference lines. Events lying to the right of the diagonal line ($\sim$90\%) are rejected. The putative fusion events cluster at low F deviation and high O deviation.}
\label{fig:OFDev}
\end{center}
\end{figure}

Fig.~\ref{fig:OFDev} shows the deviation of each event at this stage from O and F reference lines.  Most of the events exhibit a smaller deviation from O than F. Well separated from most of the events is a small group that manifests a smaller deviation from F than from O, hence are putative proton fusion events. 
The diagonal dashed (red) line in Fig.~\ref{fig:OFDev} separates events into those that are more \enquote{O-like} or \enquote{F-like}. 
By selecting traces that only lie above the dashed (red) line $\sim$ 90\% of the remaining 
beam-like events are rejected. 
From Fig.~\ref{fig:OFDev} it is clear that some beam-like events which lie close to the diagonal line are retained. These events are rejected as subsequently described. 

The location at which the proton fusion occurs was determined by comparing E$_A$ with an anode-dependent threshold, E$_{th}$. These thresholds were determined by fitting the beam E$_A$ distribution for each anode with a Gaussian. The Gaussian centroids were then fit with a second-order polynomial depicted as the average beam line in Fig.~\ref{fig:OFAvg}. The threshold, E$_{th}$, was then determined by adding 190 keV ($\sim$3$\sigma$) to the beam average in order to account for the dispersion (FWHM) in energy deposit of the beam. 

\begin{figure}[h]
\begin{center}
\includegraphics[width=0.35\textwidth]{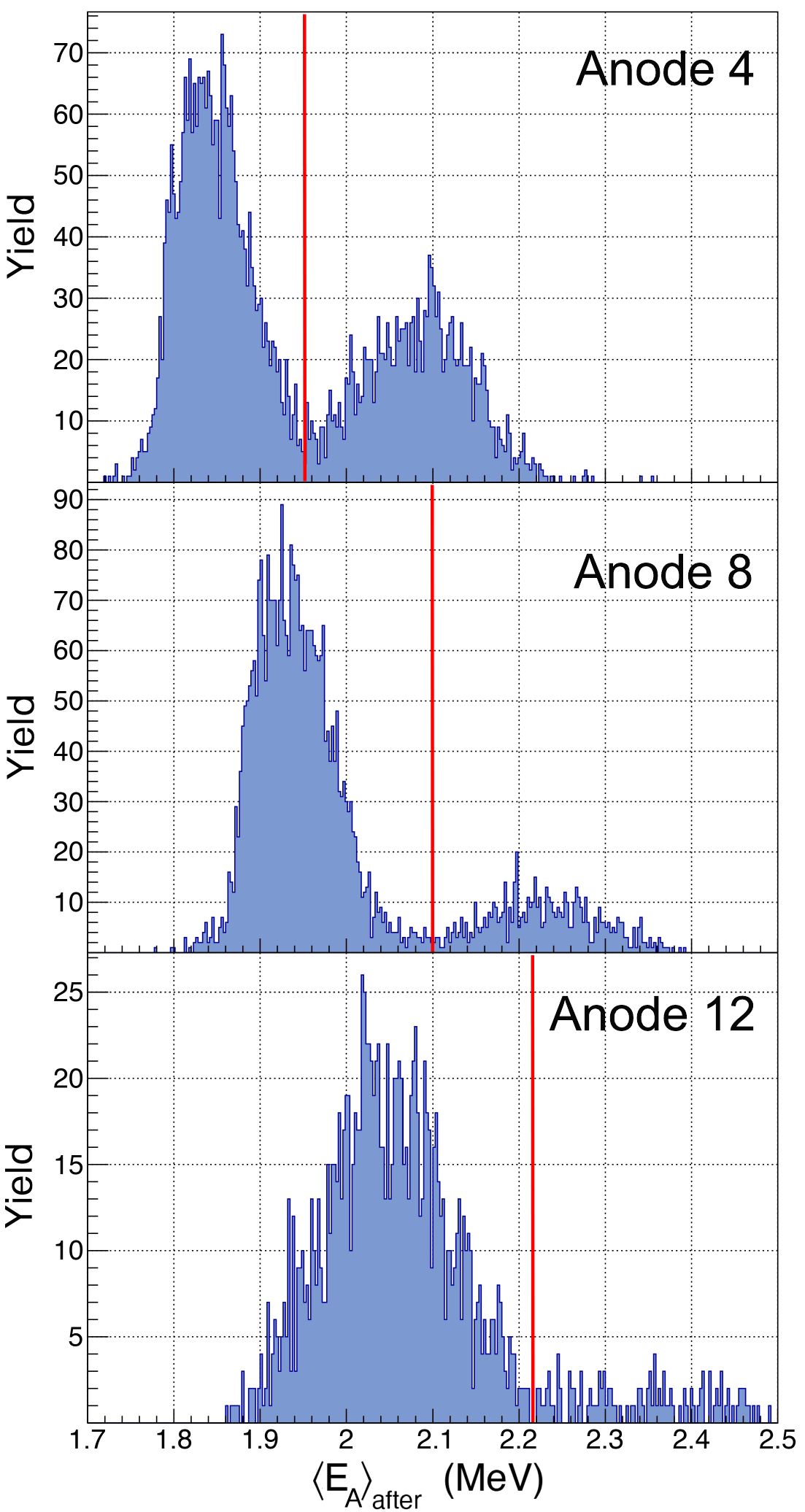}
\caption{Distribution of $\langle$E$_{A}\rangle$$_{after}$. Two peaks are visible: the left peak corresponds to scattered-beam events, and the right peak corresponds to proton-fusion events. The beam-scattered events are rejected by a threshold indicated by the vertical (red) line.}
\label{fig:Eafter}
\end{center}
\end{figure}

To further select proton fusion, specifically isolating it from proton two-body scattering (elastic or inelastic), the quantity $\langle$E$_{A}\rangle$$_{after}$ was constructed.
The quantity $\langle$E$_{A}\rangle$$_{after}$ is the average energy deposit from the anode where E$_{th}$ is exceeded up to the anode where E$_{A}^{max}$ occurs. The distribution of $\langle$E$_{A}\rangle$$_{after}$ for anodes 4, 8 and 12 is presented in Fig.~\ref{fig:Eafter}.
In all cases, a two-peaked distribution is evident. To determine the nature of the events associated with each peak, individual traces were examined after selecting on either the lower or higher peak in $\langle$E$_{A}\rangle$$_{after}$. Traces associated with the lower peak manifested the characteristic energy deposit of two-body scattering from a proton, while traces associated with the larger $\langle$E$_{A}\rangle$$_{after}$
were associated with proton fusion. The vertical (red) line indicates the minimum value used to select proton fusion events.
The bimodal nature of the  $\langle$E$_{A}\rangle$$_{after}$ distribution is clearly evident for anodes prior to 12 as evident in Fig.~\ref{fig:Eafter}, allowing selection of proton fusion for these anodes. 
The loss of separation between these two reactions in $\langle$E$_{A}\rangle$$_{after}$ beyond anode 12 for $^{20}$O and anode 10 in $^{19}$O provided the lowest incident energy for which the cross section was determined.

\section{Determination of the proton fusion cross section}

Having isolated proton fusion events, $N_{ER}$, from beam and other reaction types, the cross section was determined. The fusion cross section, $\sigma_F$ is related to the number of proton fusion events, the number of incident beam ions, $N_{Beam}$, and the thickness of an anode strip, $\Delta$x by: $\sigma_F = N_{ER}/(N_{Beam}\times\Delta x)$. 
Complete detection of the ER within the active volume of MuSIC@Indiana eliminates the need for an efficiency correction. Unlike thin-target measurements, which are sensitive to the measured ER angular distribution,  MuSIC measurements intrinsically provide an angle-integrated measurement of the cross section. The uncertainty in the measured cross section is largely determined by the statistical uncertainty associated with $N_{ER}$. 
In addition, a systematic uncertainty of 5$\%$ associated with the possible misidentification of fusion events for $N_{ER}$ is included.  The uncertainty associated with $N_{Beam}$ is negligible, and the uncertainty associated with $\Delta x$ is defined by the pressure variation in MuSIC@Indiana. All these uncertainties are included in the reported error bars.
The finite size of each anode strip results in an uncertainty in the energy at which the reaction occurs.

\begin{figure}
\begin{center}
\includegraphics[width=0.45\textwidth]{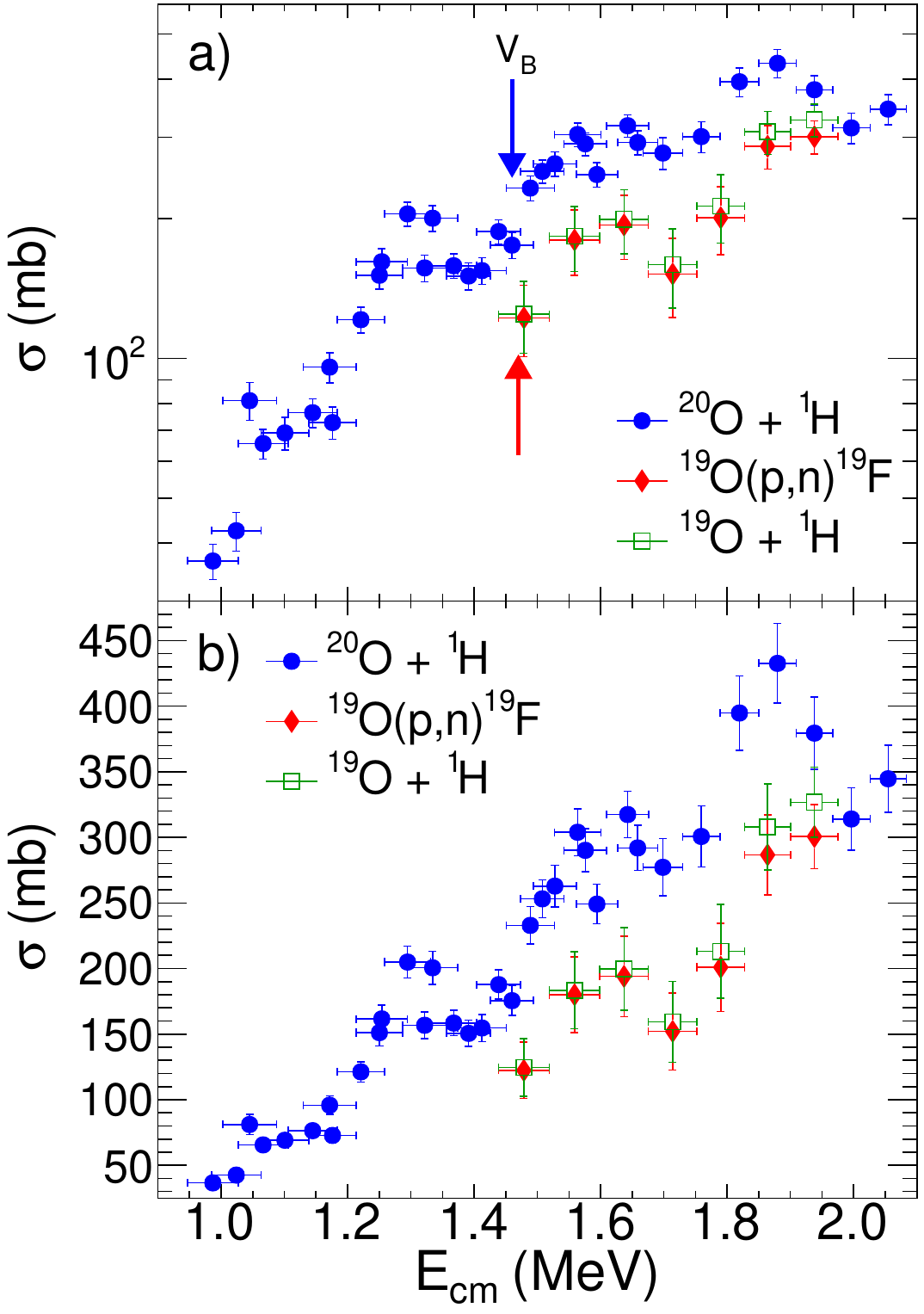}
\caption{Dependence of the cross section for $^{19}$O(p,n) and $^{20}$O(p,n) on E$_{cm}$ in a) log and b) linear scale. The Bass barrier, V$_{B}$ \cite{Bass74} for each system is indicated by the arrows: V$_{B}$ = 1.47 MeV for $^{19}$O and V$_{B}$ = 1.46 MeV for $^{20}$O.}
\label{fig:XsectErr}
\end{center}
\end{figure}

In Fig.~\ref{fig:XsectErr} the measured excitation functions for $^{19}$O and $^{20}$O are presented. The excitation function for $^{20}$O is measured over the interval 1.0 $\leq$ E$_{cm}$ $\leq$ 2.05 MeV, reaching a cross section level of $\sim$40 mb at the lowest energy. In the case of $^{19}$O a smaller interval, 1.48 $\leq$ E$_{cm}$ $\leq$ 1.94 MeV, was measured. 
Consistent with a barrier-governed process, one observes a general decrease in the cross-section, $\sigma$, with decreasing E$_{cm}$.

For $^{20}$O a non-smooth behavior is observed with an indication of three broad peaks superimposed on the overall excitation function. 
This observation was first reported in \cite{Desilets26}. As the implications of the oscillatory structure were previously described in detail, they are only briefly summarized below.

This oscillatory behavior of the proton fusion cross section is interpreted as the formation of one or more short-lived quasi-bound states which subsequently decay by emission of a single neutron. At the low incident energy, only low angular momentum states (J = 1/2, and possibly 3/2) are accessible. Thus, not the entire level density of $^{21}$F at this excitation is probed; only the level density of low-angular-momentum states is relevant. On general grounds, the oscillatory behavior is impacted by the number and spacing of the quasi-bound states, the level density of states in the $^{20}$F following neutron emission, and the coupling between the quasi-bound and $^{20}$F states.  

These peaks are reminiscent of broad peaks observed for stable nuclei. For an $^{18}$O target, broad peaks were measured in (p,$\gamma$) \cite{Butler59}, (p,n) \cite{Hill56}, and (p,$\alpha$) \cite{Hill56, Clarke59, Carlson61,Lorenz-Wirzba79}. Peaks in the differential cross-section spectrum were described as resonant states in the compound $^{19}$F nucleus. 

In light nuclei, this observation of broad peaks may suggest that the standard statistical approach to level density, namely a monotonically increasing function with excitation energy, and level properties near the barrier, is likely inadequate \cite{Kumar25}. In $^{21}$F at an excitation of $\sim$9 MeV, observed individual levels are separated by $\sim$300-600 keV \cite{NuDat3}. At the higher excitation energy of this work, E$^*$$\sim$13 MeV, a separation of $\sim$200 keV is not unreasonable.

In the case of $^{19}$O, despite the statistical quality of the data, two features are immediately clear. Overall, a significant suppression of the cross section is evident relative to $^{20}$O. In addition, as in the case of $^{20}$O a non-smooth behavior with energy is observed. The slight suppression in $\sigma$ for $^{20}$O at E$_{cm}$$\sim$1.7 MeV is also reflected in a suppression of $\sigma$ for $^{19}$O at approximately the same energy. Unfortunately, the limited energy range spanned for $^{19}$O does not permit determination of the dip at E$_{cm}$$\sim$1.4 MeV observed in the case of $^{20}$O. 

\begin{figure}
\begin{center}
\includegraphics[width=0.45\textwidth]{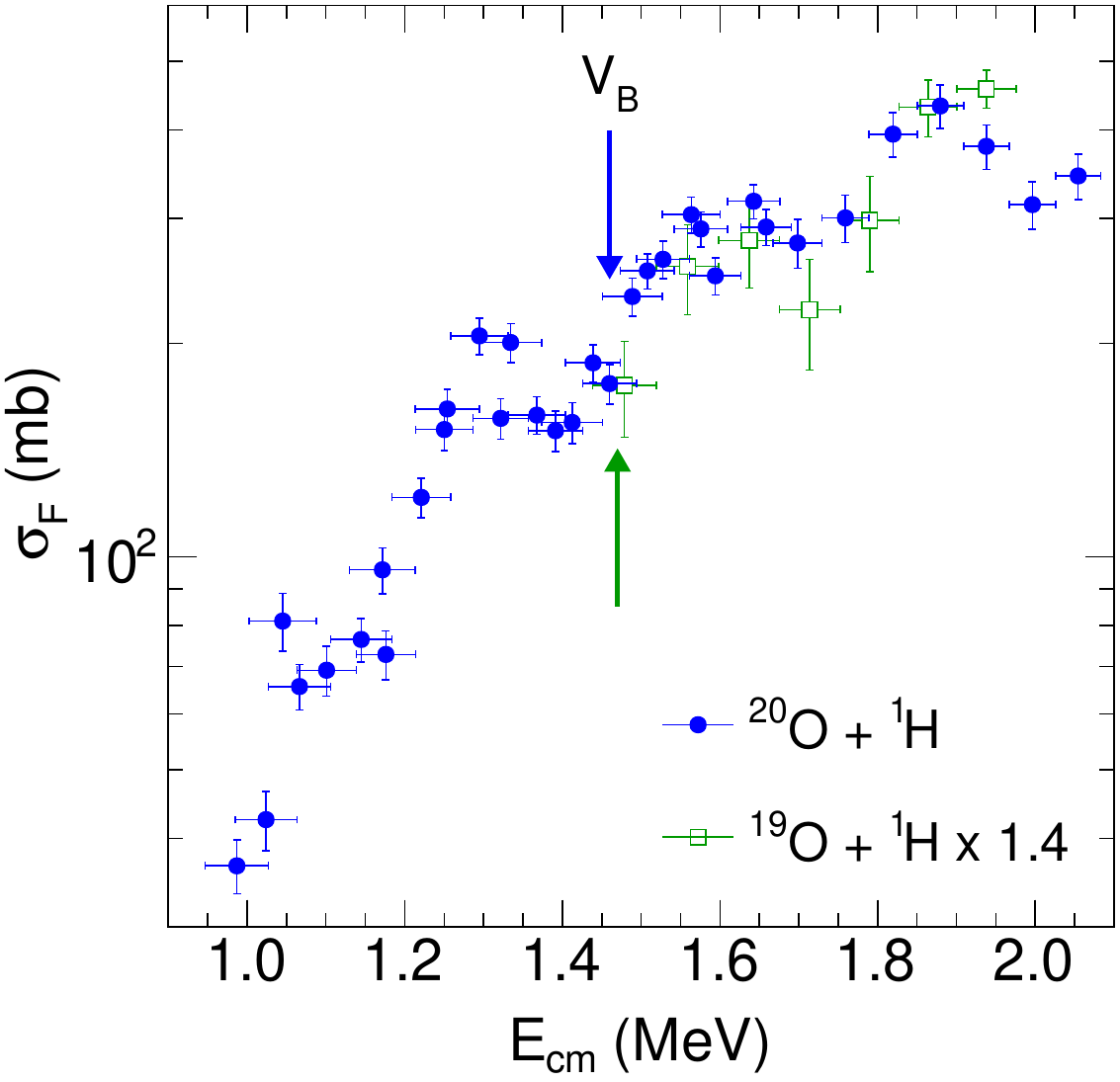}
\caption{Comparison of the total fusion cross section for $^{20}$O (filled circles, blue) with the extracted fusion cross section for $^{19}$O scaled by a factor of 1.4 (open squares, green). For reference, the Bass fusion barrier is indicated by the arrows.}
\label{fig:XsectErr_v2}
\end{center}
\end{figure}

Presented in Table\ref{table:decay_chan} are the de-excitation channels calculated following fusion for $^{19,20}$O in the Hauser-Feshbach statistical decay model code GEMINI++ \cite{Charity10}. For $^{20}$O, single-neutron emission accounts for $>$99.8$\%$ of the cross section, while for $^{19}$O some $\alpha$-emission reduces single-neutron emission to as low as 92$\%$ at the highest measured energies. Thus, for $^{20}$O, the (p,n) cross section measurement corresponds directly to the fusion cross section. Because $^{19}$O has other decay channels, extracting the fusion cross section requires either accounting for their influence or measuring them directly. 

In the case of $^{19}$O, the total fusion cross section is determined by scaling the measured (p,n) cross section by the relative yield predicted by the GEMINI++ calculations in Table\ref{table:decay_chan19}. This total fusion cross section is represented by the open squares (green) in Fig.~\ref{fig:XsectErr}. Clearly, the proton fusion cross section for $^{19}$O is significantly less than that of $^{20}$O.

To facilitate comparison of the structure evident in the proton fusion excitation functions of $^{19}$O and $^{20}$O,  the fusion cross section for $^{19}$O was scaled by a factor of 1.4 as presented in
Fig.~\ref{fig:XsectErr_v2}.
The scaled cross section for $^{19}$O agrees remarkably well with $^{20}$O in both the location and magnitude of the oscillatory behavior. 

To understand why $^{19}$O exhibits a significantly smaller proton fusion cross section, it is necessary to consider other channels that might open and reduce the flux resulting in fusion. For $^{19}$O with its unpaired neutron, the single-neutron transfer Q=-1.73 MeV. In contrast, for single neutron transfer in $^{20}$O, Q=-5.38 MeV. Thus, one might expect that when E$_{cm}$ exceeds 1.73 MeV, the fusion cross-section is reduced for $^{19}$O but not for $^{20}$O. The unfavorability of two-neutron transfer for $^{20}$O below its Q-value of -3.08 MeV eliminates two-neutron transfer in the energy range examined. Fig.~\ref{fig:XsectErr} shows that the fusion cross section is suppressed even at E$_{cm}$=1.5 MeV, the lowest energy measured, indicating that single-neutron transfer does not play a major role in the fusion suppression. 

An alternate explanation for the suppression of the fusion cross section in $^{19}$O as compared to $^{20}$O may lie in consideration of
the relative level densities of both the quasi-bound compound nucleus as well as the daughter evaporation residue formed following single neutron emission.
As the change in atomic number for proton capture of both $^{19}$O and $^{20}$O is the same, the neutron level density is the primary consideration.

For $^{19}$O, the resulting evaporation residue, $^{19}$F has an even number of neutrons while 
for $^{20}$O, the resulting evaporation residue, $^{20}$F has an odd number of neutrons. Thus, based on the evaporation-residue neutron number, even at the same excitation, one would expect a higher level density for $^{20}$F. This expectation is compatible with the higher observed cross section for $^{20}$O. This suggests that any difference in the level densities of the quasi-bound compound nuclei is overcome by the level densities of the resulting evaporation residues.

In principle, extending this line of reasoning opens up an interesting possibility. If the level density, or relative level density, of the decay stage could be calculated (or measured), then the level density (or relative level density) of the quasi-bound compound nucleus could be extracted. Such information could be valuable for validating the predictions of theoretical models such as the Gammow shell model in the coupled channel representation --the appropriate tool to describe near-threshold spectroscopy and reactions \cite{Jaga14,Michel21,Fernandez23}. 

\begin{table} [t!]

  \begin{center}
    \begin{tabular}{cccccc}
    \hline
    {CN} & {E$^*$ (MeV)} & {n(\%)} & {$\alpha$(\%)} &  {$\alpha$n(\%)} & {p(\%)}\\
    \hline
    $^{20}$F & 12.1 & 98.2 & 1.8 & 0.0 & 0.0 \\
    $^{20}$F & 12.6 & 92.0 & 2.3 & 5.7 & 0.0 \\
    $^{21}$F & 12.2 & $>$99.9 & 0.0 & 0.0 & 0.0 \\
    $^{21}$F & 13.3 & 99.8 & 0.1 & 0.0 & 0.1 \\
    
   \bottomrule
    \end{tabular}
  \end{center}
  \caption{GEMINI++ \cite{Charity10} predictions for the yield of emitted particles from $^{20}$F and $^{21}$F nuclei formed in fusion of $^{19,20}$O with proton.}
  
  
  \label{table:decay_chan} 
\end{table}

\begin{table} [t!]

  \begin{center}
    \begin{tabular}{cccccc}
    \hline
    {E$_{cm}$ (MeV)} & {E$^*$ (MeV)} & {n(\%)} & {$\alpha$(\%)} &  {$\alpha$n(\%)} & {p(\%)}\\
    \hline
    1.48 & 12.12 & 98.20 & 1.80 & 0.00 & 0.00 \\
    1.56 & 12.20 & 98.15 & 1.85 & 0.00 & 0.00 \\
    1.64 & 12.28 & 97.20 & 1.73 & 1.00 & 0.07 \\
    1.71 & 12.35 & 95.40 & 2.34 & 2.17 & 0.09 \\
    1.79 & 12.43 & 94.34 & 1.87 & 3.67 & 0.12 \\
    1.86 & 12.50 & 93.08 & 2.09 & 4.79 & 0.04 \\
    1.94 & 12.58 & 92.00 & 2.26 & 5.67 & 0.07 \\
    
   \bottomrule
    \end{tabular}
  \end{center}
   \caption{Yield of decay channels predicted by GEMINI++ \cite{Charity10} following proton fusion on $^{19}$O.}
  
  \label{table:decay_chan19} 
\end{table}

\section{Comparison with theory}
To better understand the measured proton-fusion cross sections, we have performed coupled channel (CC) calculations. The theoretical calculations were performed using the FRESCO code~\cite{THOMPSON1988167}. For the real part of the optical potential, the São Paulo potential (SPP)~\cite{CPH97} was used. This potential includes a systematic dependence on matter density based on experimental electron scattering and Dirac Hartree-Bogoliubov calculations. The same systematic was used for the $^{20}$O~+~p system~\cite{Desilets26}. For the imaginary part, a short-range Woods-Saxon potential, W$_{\rm{F}}$(r), was used for both nuclei with the parameters $\rm{W_{0}=50.0}$~MeV, $\rm{r_{W}=1.06}$~fm, and $\rm{a_{W}=0.2}$~fm for the depth, reduced radius, and diffuseness, respectively. This short-range potential exclusively accounts for absorption to fusion~\cite{CDH18}. 

The 1.4717 MeV $1/2^+$ excited state was considered in the CC calculations, as this is the only state for which a considerable quadrupole reduced transition probability (B(E2) = 0.58 W.u.) is reported~\cite{NuDat3}. We therefore used a model-independent coupling based on this B(E2) value. We also considered other low-lying excited states up to the 5/2$^{+}$ state at 3.1535 MeV in the coupling scheme, assuming the same B(E2) for the other states. However, only very weak reduced dipole magnetic transition probabilities are reported for the states up to the 9/2$^{+}$ state at 2.3710 MeV. The fusion cross section results remain unchanged over the studied energy interval when considering all these states.

Recall that the $^{19}$O+p system is studied at center-of-mass energies between 1.5 MeV and 2 MeV.
Once the excitation energies of the $^{19}$O projectile are very close to the threshold of the $1/2^+$ excited state, the S-matrix solution may present numerical issues. In this regard, one can use the R-matrix method to compute the observables~\cite{Descouvemont_2010}.
The calculations converged with only N$_{\rm{max}} = 30$ basis states, and a matching radius of R$_{\rm{match}} = 15$~fm.

It is important to note that the default option for the fusion cross section calculation for a given total angular momentum $J$ in the FRESCO code is

\begin{equation}
    \sigma_{\rm{F}} (J)= \sigma_{\rm{R}} (J) - \sum_\alpha \sigma_{\rm{dir}}^\alpha (J), 
\label {sf-diff}
\end{equation}
where $\sigma_{\rm{dir}}^\alpha (J)$ is the cross section of the direct reaction channel $\alpha$ (inelastic excitation of $^{19}$O projectile, in this case). In Eq.~\ref{sf-diff}, $\sigma_{\rm{R}} (J)$ is the partial reaction cross section derived from the R-matrix solution for the elastic channel~\cite{thompson2009nuclear}.
For large $J$ values, both terms in the real part of Eq.~\ref{sf-diff} are very small, and due to numerical inaccuracies, the code even prints negative partial fusion cross sections. These inaccuracies can be addressed by using the following expression for the calculation of $\sigma_{\rm{F}}$

\begin{equation}
    \sigma_{\rm{F}}= \ \frac{\pi }{K^2}\sum _\alpha \left< \Psi _\alpha\left |W_{\rm{F}} \right |\Psi _\alpha\right>, 
\label {sf-integral}
\end{equation} 
where $\Psi _\alpha$ is the wave function of the channel $\alpha$ and $K$ is the wave number~\cite{canto2013scattering}.

\begin{figure}
\begin{center}
\includegraphics[width=0.45\textwidth]{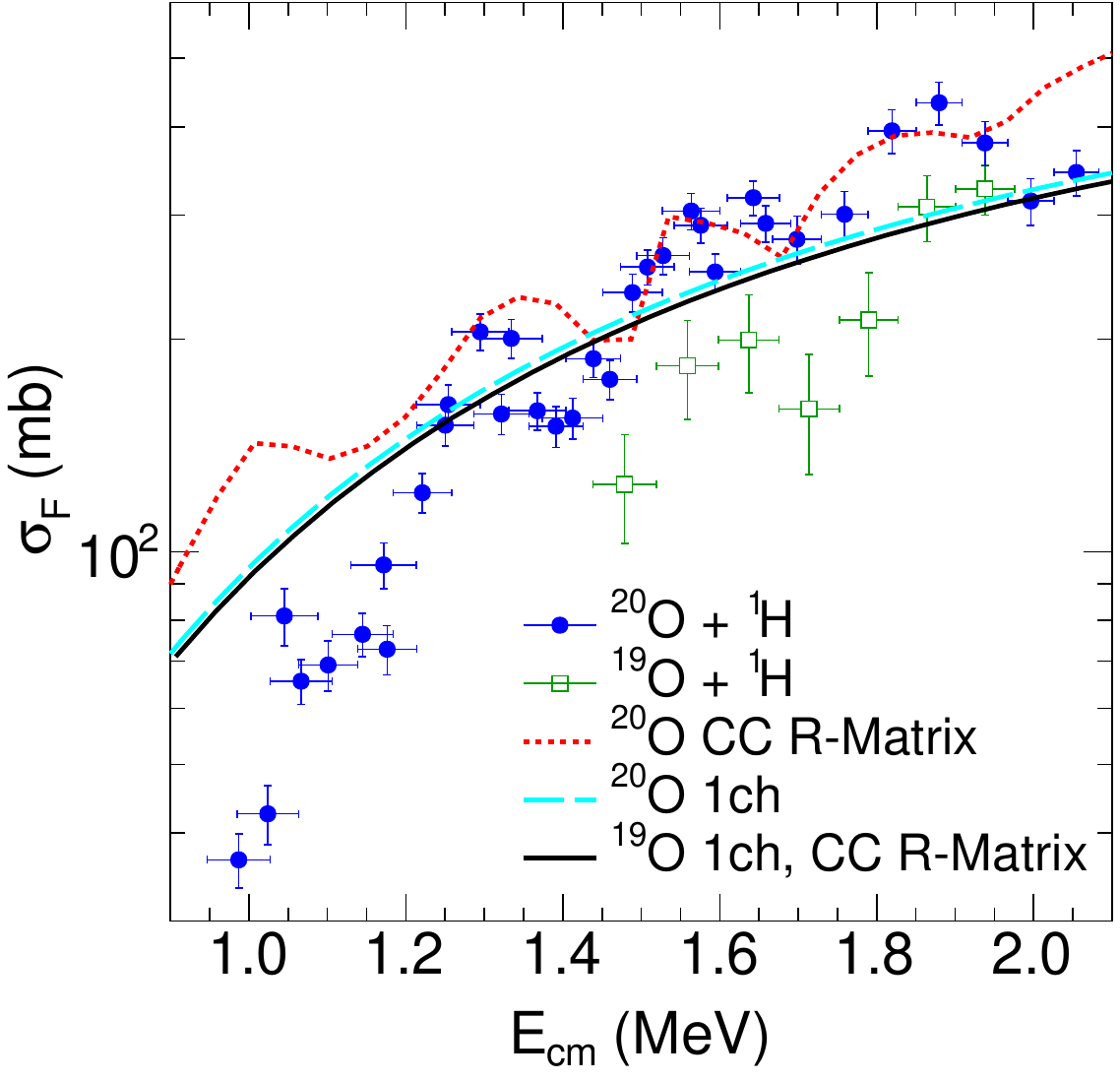}

\caption{Fusion cross section for the reaction $^{20}$O + $^{1}$H and $^{19}$O + $^{1}$H. The predicted cross section from one channel and R-matrix coupled channel calculations for $^{20}$O + $^{1}$H is also depicted.}
\label{fig:XsectErr_CC}
\end{center}
\end{figure}

The experimentally measured fusion excitation functions are compared in Fig.~\ref{fig:XsectErr_CC} with coupled channel calculations.
The theoretical predictions designated as one-channel correspond to the interaction of the ground-state density distributions for the oxygen and hydrogen nuclei. Only a minimal difference is observed between $^{19}$O and $^{20}$O at the highest energies. In the same figure, we compare the one-channel calculation with the CC calculations. One observes that couplings to the inelastic excitation of the target do not affect the fusion cross section of the $^{19}$O+p system at near-barrier energies (V$_{\rm{B}}$ = 1.46 MeV) in contrast with the results previously observed for the $^{20}$O+p system \cite{Desilets26}. As the one-channel and CC results coincide in the studied energy interval, both are represented by the full line in Fig~\ref{fig:XsectErr_CC}.  This may reflect differences in the level densities of the compound and residual nuclei after neutron evaporation in each reaction. Nevertheless, the experimental fusion cross section for the $^{19}$O+p system shows oscillatory behavior similar to that observed for the $^{20}$O+p system. The theoretical cross section also overestimates the mean experimental value. We could not find any explanation for that within the theoretical model used to derive the fusion cross section for the system studied in this paper. This may encourage researchers to seek an explanation for this unresolved issue.

Finally, to seek any possible reason for the disagreement between theoretical and experimental fusion cross sections for the $^{19}$O+p system, we assumed the B(E2:$2^+ \to 0^+$ ) value for $^{20}$O in the B(E2:$1/2^+ \to 5/2^+$ ) for $^{19}$O, but the CC result for the fusion cross section remained unchanged.

\section{Conclusions}
Use of the active target detector MuSIC@Indiana with CH$_{4}$ gas and incident radioactive beams of $^{19}$O and $^{20}$O ions 
enabled efficient extraction of the (p,n) cross section. This approach could be readily applied to investigate (p,n) reactions for other neutron-rich light heavy-ion beams.

At the energies considered, where the role of direct reactions is negligible, this measurement enables extraction of the proton fusion cross-section, $\sigma_F$, due to the dominant decay of the compound nucleus via emission of a single neutron. Two main experimental facts emerge from this work.
As previously observed in the case of $^{20}$O, proton fusion in $^{19}$O manifests a clear oscillatory behavior. In addition, the overall cross-section for $^{19}$O is observed to be  suppressed relative to $^{20}$O by $\sim$30\%. Scaling the $^{19}$O excitation function demonstrates that the location and magnitude of the oscillations in both cases are quite similar suggesting a common underlying origin. 
The suppression of the fusion cross section is hypothesized to be due to a reduced neutron level density in the evaporation residue $^{19}$F as compared with the neutron level density of $^{20}$F. 

CC calculations for the $^{19}$O + p system using R-matrix theory could not reproduce the oscillatory behavior of the fusion excitation function observed experimentally. Moreover, the theoretical cross section also exceeds the mean experimental fusion cross section. 

To provide a better and comprehensive understanding of proton fusion in such light systems further effort on both experimental and theoretical fronts is clearly necessary.

\section{Acknowledgments}
We acknowledge the high-quality beam and experimental support provided by the technical and scientific staff at the Grand Acc\'{e}l\'{e}rateur National d'Ions Lourds (GANIL), in particular D. Allal, B. Jacquot, and D. Gruyer. 
We are thankful for the high-quality services of the Mechanical Instrument Services and Electronic Instrument Services facilities at Indiana University.

This work was supported by the U.S. Department of Energy Office of Science under Grant No. 
DE-SC0025230 (IU) and DE-SC0021938 (MSU), the National Science Foundation under Grant No. PHY-2309923 (MSU), and Indiana University. This research has received funding from the European Union's HORIZON EUROPE Program under grant agreement no. 101057511. R. deSouza gratefully acknowledges the support of the GANIL Visiting Scientist Program. Kyle Brown also acknowledges support from Michigan State University.

Brazilian authors were partially funded by CNPq, FAPERJ, and the INCT-FNA (Instituto Nacional de Ci\^encia e 
Tecnologia-F\'\i sica Nuclear e Aplica\c c\~oes), research project 408419/2024-5.

 \textit{Data Availability Statements}. 
The supporting data for this article are from the e831\_{21} experiment and are registered as
https://doi.org/10.26143/GANIL-2023-E831\_21 following the GANIL Data Policy.




\bibliographystyle{elsarticle-num} 






\end{document}